\documentclass{llncs}
\usepackage[cmex10]{amsmath}
\usepackage{siunitx}
\usepackage{array}
\usepackage{mdwtab}
\usepackage{hyperref}
\usepackage{booktabs}
\usepackage[final]{graphicx}
\usepackage{todonotes}
\usepackage{array}
\usepackage{makecell}

\newcolumntype{L}{>{\centering\arraybackslash}m{2.0 cm}}

\begin{document}
\title{A Case for a Currencyless Economy Based on Bartering with  Smart Contracts}

%

\author{Carlos Molina--Jimenez\inst{1}
  \and Hazem Danny Al Nakib\inst{2}
  \and Linmao Song\inst{3}
  \and Ioannis Sfyrakis\inst{4}
  \and Jon Crowcroft\inst{1}
 } 

\institute{
Department of Computer Science and 
Technology, University of Cambridge\\
\email{carlos.molina@cl.cam.ac.uk}
\url{https://www.cl.cam.ac.uk/~cm770/}
\email{jon.crowcroft@cl.cam.ac.uk}
\and
Department of Computer Science, University College London \\
\email{hazemdanny.nakib@gmail.com}\\ 
\and 
Independent researcher \\
\email{linmao.song@gmail.com}
\and
School of Computing, Newcastle University, UK\\
\email{ioannis.sfyrakis@ncl.ac.uk}
}%

\date{\today}  

\maketitle              

\noindent
\makebox[\linewidth]{\small University of Cambridge, 8 Oct 2020}

\begin{abstract}
We suggest the re--introduction
of bartering to create a cryptocurrencyless,
currencyless, and moneyless economy segment.
We contend that a barter economy would benefit
enterprises, individuals, governments and societies.
For instance, the availability of an online peer--to--peer
barter marketplace would convert ordinary individuals
into potential traders of both tangible and
digital items and services. For example, they will be able
to barter files and data that they collect. Equally
motivating, they will be able to barter and re--introduce
to the economy items that they no longer need 
such as, books, garden tools, and bikes which
are normally kept and wasted in garages and sheds.
We argue that most of the pieces of technology needed
for building a barter system are now 
available, including blockchains, smart contracts,
cryptography, secure multiparty computations and fair 
exchange protocols. However, additional research is
needed to refine and integrate the pieces together.
We discuss potential research directions. 

\end{abstract}

\section{Introduction}
\label{introduction}
In this article, we make a case for re--introducing the barter
trade model, underpinned by the latest advances in decentralised
technologies. We focus mainly on peer--to--peer (direct) bartering
as opposed to indirect bartering where the barter operations 
(also called transactions) are  mediated by a trade exchange.
Such a system would allow the execution of both face--to--face 
and remote barter transactions to trade both tangible and 
digital items, and services. To this end, bartering would target a 
specific economic segment and complement the current trade models.
We present our views on the limitations of the current 
trade systems, the potential of a barter system and the 
challenges involved.

\begin{figure}
\centering
\includegraphics[width=0.55\columnwidth]{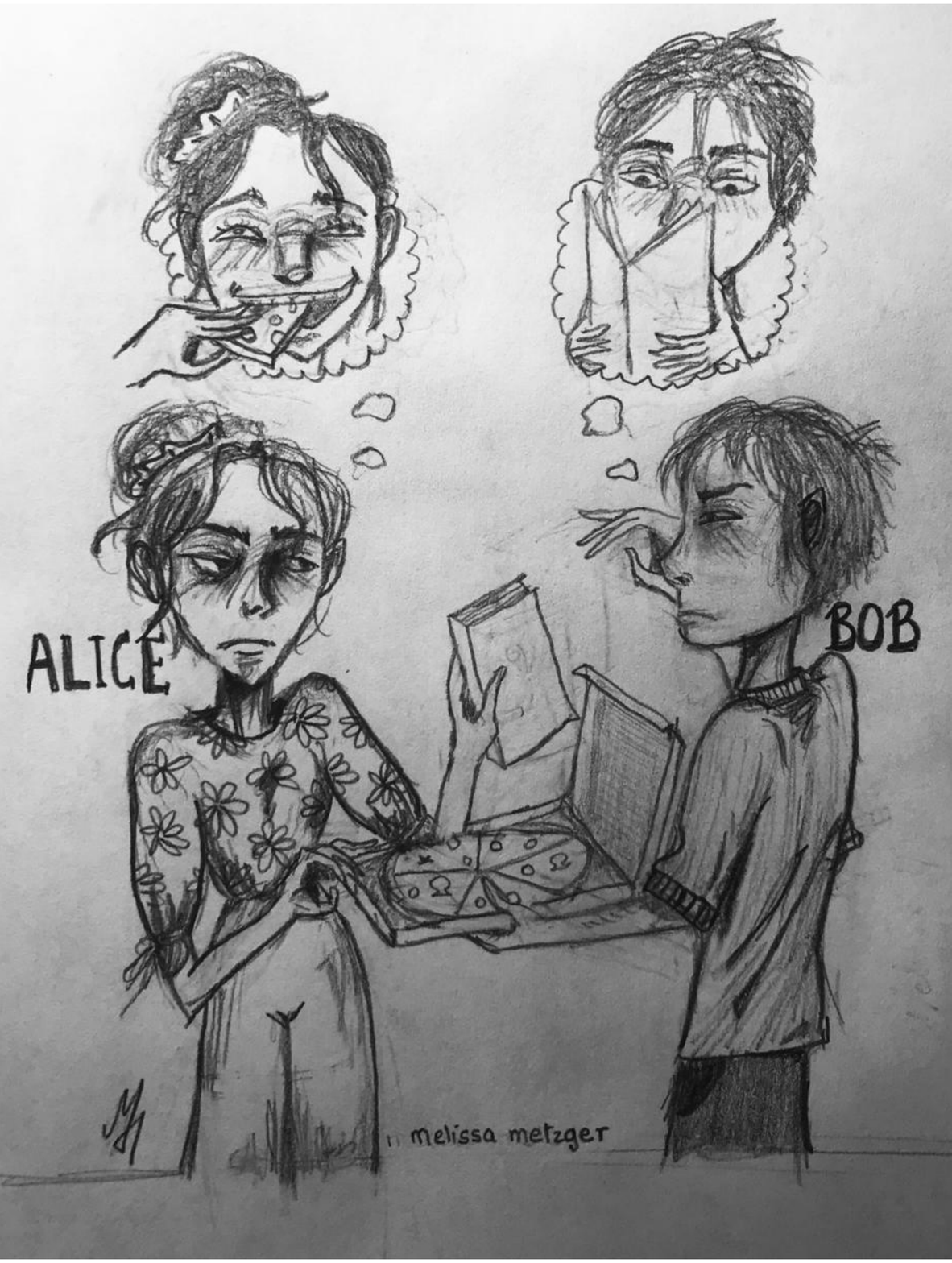}
\caption{Alice and Bob bartering tangible items.}
\label{fig:aliceandbobbartering}
\end{figure}

The most attractive features of bartering is that it is
inherently peer--to--peer and bank--independent: it consists in
the exchange of item--for--item on a peer--to--peer basis. 
Fig.~\ref{fig:aliceandbobbartering} illustrates the idea: 

Alice meets Bob and she gives her book  to him in exchange for 
a pizza. In this, and subsequent figures, we use Alice and Bob  
to represent two participants in a barter transaction. 
In some situations, Alice and Bob are humans
operating on their own initiative but in others they represent 
institutions like companies and governments. Their interaction is
either face--to--face or remote, that is, online.

Bartering was widely practiced in all eras until 
it was replaced by the currency--based trade model that 
underpins today's trade through both  private currencies and 
state-backed government led fiat monies~\cite{IntuitInc2018}.  

Why do we care about bartering now?  We are interested in bartering
because it has intrinsic features that in some situations make it 
more suitable than the current fiat--money--based trade system. 
We care about it now because, after decades of research we now 
have all the pieces of technology to implement it, admittedly, after 
solving some pending research questions. 

The remaining sections of this article are organised as follows:
Section~\ref{currenttradelimitations} includes background
material to help understand the advantages and disadvantages
of the current fiat--based trade system. This is followed
by Section~\ref{motivatingexamples} which introduces
bartering and discusses its potential contribution to the economy.
Section~\ref{researchchallenges} outlines research 
questions that need to be solved to progress towards
a potential implementation of a peer--to--peer barter system 
that can complement the existing trade system. In Section~\ref{furtherread}, 
we suggest references where the topics outlined 
in this article are discussed in depth.
Concluding remarks are presented in Section~\ref{conclusions}.

\section{Pros and cons of fiat--based trade}
\label{currenttradelimitations}
The origin of currency--based trade has a rich history that can be traced back to as early as circa 9000--6000 BC. There is evidence that cattle were used as a 
medium of exchange, that is, as money~\cite{RoyDavies2020}. Different 
societies used different objects such as beads, shells, furs, seeds and
metals. These objects were gradually replaced by coins, minted coins and finally 
paper money~\cite{Paul1947}. Private moneys were used until governments introduced fiat money like dollars and euros.

Currently, in the post-Gold Standard era, fiat currencies that are not backed
by any assets dominate trade, however, cryptocurrencies like Bitcoin 
are emerging. Bartering predates the appearance of currency--based trade 
and operated fairly actively alongside the currency--based model until the appearance of market--integrated economies. In such economies, currency offers noticeable advantages over bartering; for instance, it helps
to maintain consistency within marketplaces, it provides efficiency 
and scalability, and simple access and ease of use.

\subsection{Advantages}  
\label{moneybasedtradedavantages}
Currency is a common medium of exchange that helps individuals to 
exchange items and services indirectly. We are not very interested in the concept of money and the requirements of money per se but rather in how and in what 
way currency compares with barter models. Relevant to our discussion are the
following properties of currency:

\begin{itemize}
    \item eliminates the difficulties of finding a matching barterer 
          to perform a simultaneous tit for tat barter operation. 
    \item is a common and standard measure of value.
    \item is a medium of condensing and accumulating wealth.
\end{itemize}    

These properties of currency significantly simplify trade operations.

\subsection{Disadvantages}
A salient feature of currency--based trade is that it \textbf{is inherently centralised}: the bank plays the role of a trusted third party that is
responsible for mediating all the trade operations.
From this fact, emerge the strengths discussed above in Section~\ref{moneybasedtradedavantages} and the weaknesses 
pointed out by some authors~\cite{Trent2016,Joseph2016}
and summarised as follows:

\begin{itemize}
  \item \textbf{mediated trade:} fiat money is inherently 
    a mediation mechanism and as such it prevents peer--to--peer trade 
    as conducted in the example of Fig.~\ref{fig:aliceandbobbartering}.
  \item \textbf{transparency:} the mediating role of the banking system 
    compromises transparency. Transaction records are regarded as private
    data that only the bank can access.
  \item \textbf{privacy:}  nothing escapes from the banker's sight. The bank 
     gathers records of absolutely all transactions.
     Cash payment offers better privacy protection, however,
     with the success of online shopping, the use of cash is becoming old 
     fashioned, in particular with the threat of the SARS--CoV--2 virus.
  \item \textbf{delays:} bank transactions are still unacceptably slow; for 
     example wire payment can take up to five days to complete.
     Non surprisingly, payment and money transfer mechanisms like PayPal, 
     Western Union and TransferWise have emerge to 
     help reduce payment delays to minutes or seconds; they
     are based on the execution of digital accounting operations that
     conceal, from the customer's view, the bank delays. 
     
  \item \textbf{fees:} banks services are not free. The charges discourage
         the execution of transactions of small value. Moreover, banks 
         lend out most of the capital, in fact banks are indebted to their 
         depositors, a point that many people misunderstand.
   \item \textbf{financial exclusion:} bank services are granted only to
     those that satisfy the bank's requirements which are normally related 
          to age, financial status, legal status and so on. 
          It has been estimated that within the current 
     world population of 7.8 billion (\num{7.8e9}), about 1.7 billion of adults remain unbanked. The reason being that they fail to meet the requirements or that there are no banking services where they live~\cite{WorldBank2017}. 
     This disappointing figure includes 1.5 million unbanked in Great Britain and 40 million in the European Union notwithstanding the fact that within the European Union it is a requirement for financial institutions to provide free bank accounts to those who need them.
\end{itemize}


As explained in Section~\ref{motivatingexamples}, the decentralised 
(peer--to--peer) nature of bartering and its non--reliance on
banks makes it immune to the drawbacks listed above. 
Bartering  can be used as an option, in particular where
the advantages that currency offers are irrelevant to the traders, 
for example, where the difficulty to find  barter partners is not an 
issue, say, because the Internet offers mechanisms to surmount the obstacle.

Recently introduced cryptocurrencies which are based on
decentralised technologies (for example, blockchains) have generated a 
lot of excitement in finance and other fields. Though they can be used
to exchange tokens, we do not see their potential to support
bartering. Firstly, the ledger that mediates all cryptocurrency
transactions introduces unnecessary complexity for the needs of
barter operations. Secondly, they are focused on tokenised items. 
Thirdly, state--of--the--art cryptocurrencies are afflicted by
several drawbacks which we will discuss separately in 
Section~\ref{cryptocurrencies}.

\subsection{Observations on cryptocurrency}
\label{cryptocurrencies}
Decentralisation was one of cryptocurrencies' original goals. Despite the significant amount of interest, investment (in terms of not only money, but also, industrial and research efforts) and speculation, the real--world benefits 
of cryptocurrencies still remain to be seen, after more than a decade since their birth. One of the possible reasons for the lack of adoption in practical applications might be cryptocurrencies' built--in weaknesses. To start with, the consensus algorithms (for example, proof–of–work and proof–of–stake) that are used to synchronise the decentralised ledgers introduce efficiency and scalability  issues~\cite{Trent2016,Joseph2016}. Secondly,
the cost of cryptocurrency transactions is high. 
Bitcoin and Ethereum have already experienced average transaction fees of 54.90 and 4.15 USD, respectively~\cite{bitinfocharts2018}. There is insufficient evidence to suggest 
that cryptocurrencies are not suitable for conducting large numbers of 
transactions involving small payments. Thirdly, to the average user, cryptocurrencies are not simple to use, their focus on decentralisation
introduces some complexities. For instance, the requirement that users 
need to manage private keys introduces the risk of accidental loss. Mitigating this issue, e.g., by using exchange wallets, defeats the purpose of decentralisation and without the added benefit of protection that would come from centralised banks. 
Fourthly, another issue that afflicts cryptocurrencies is security.
A notorious example of a security hole is the MtGox case that resulted in the theft of 850,000 bitcoins (about 620 million USD~\cite{Decker2014}) in Feb 2014. Real world abuses include duplication, counterfeits, multiple--spending and money laundering~\cite{Laurie1996,Syed2015}. Cryptocurrencies also facilitates new crimes, such as ransom payments. For example, victims of CryptoLocker (a well known ransomware trojan) were forced to pay 310 472 USD in 2013~\cite{Kevin2016}.
Finally, in addition to these marketing and technical issues,
the practicality of decentralised  cryptocurrencies that allow 
individuals to conduct transactions freely outside a regulatory 
supervising bank system has been challenged by several
authors. The lack of Automated Clearing House (ACH) reversal in 
blockchain systems is only one of the concerns~\cite{KaiApr2017,KaiDec2017}.
Several mechanisms are currently under active investigation to ameliorate 
these and other issues that afflict current state-of--the--art 
cryptocurrencies.

\section{Barter--based trade}
\label{motivatingexamples}
A barter system operating along the currency--based
trade system would enable individuals and institutions
to barter a variety of items and services. 
The example of Fig.~\ref{fig:aliceandbobbartering} shows a
familiar situation: Alice gives away an item that otherwise
would have become wasted food; regarding Bob, he reintroduces 
to the economy an item that bears no value to him. To appreciate
the impact, it is worth taking into account that a study
from 2017 reports~\cite{HC2017} that the average household 
lost \pounds470 a year
because of avoidable food waste. Similarly, there are reports
that the average household in UK has 
accumulated \pounds1 784 worth of items that are no longer
needed including, books, bikes, and garden tools~\cite{Build2019}.

\subsection{Do I have something to barter?}
\label{itemstobarter}
Though the example of Fig.~\ref{fig:aliceandbobbartering}  
involves two tangible items, bartering
can be extended to cover a great variety of items and 
services. We conceive bartering as an operation
executed between the two participants. The complexity of such transaction and
the technology needed to execute it depend to a great extent
on the properties of the items. To start reasoning, it might help to distinguish between the following categories:
 
\begin{itemize}
\item \textbf{tangible items:} within this category fall physical items
  like the pizza and book of the example of Fig.~\ref{fig:aliceandbobbartering}. A salient property of these items is that they are not transferable
  over electronic communication channels. 

\item \textbf{digital items:} within this category are items that
 can be manipulated electronically, for instance, they can be stored (normally as a
 file on disk) and transferred, electronically. Examples of
 digital items are photos,  videos, e--books and pieces of
 personal data.
 Two salient properties of digital items are that they possess value in 
 their own right and that they exist only in the digital world.
 The universe of digital items is large and includes items
 with different properties and as such it can be
 further divided into other sub--categories. For example, 
 depending on their uniqueness, a digital item can be
 fungible or non--fungible~\cite{FabToken2019}.
 Bartering of digital items has   great business potential because 
 the number of digital items that each individual owns can only 
 increase and will be, in the near future, larger than the number 
 of his  or her tangible items. 

\item \textbf{online services:} an online service (also called
 e--service) is an network--accessible facility that satisfies
 some user's needs.  Their electronic nature makes them amenable
 to barter transactions. 
 As an opening example, we can mention that current access
 to online services (for example, google, gmail and facebook)
 is granted on the basis of bartering: online services exchange
 their services in return for the personal data that they
 collect about the individuals' activities. This practice has been 
 in place since the early 90s and so far it seems to satisfy both parties.
 That said, one can argue that it is an unfair barter operation
 since the economic power of the service providers allows them
 to dictate the rules.

 Another practical example of bartering of online services is 
 found in the current operation of the GBP 
 (Gateway Border Protocol). 
 A typical arrangement is a peer-to--peer relationship 
 where two ASs (Autonomous Systems) agree to exchange traffic \emph{for free}, that is, without the 
 exchange of actual money~\cite{Bruno2005}.  

  A large number of similar examples, can be found in the cloud computing domain where
 cloud providers agree to barter idle resources with each other. Another 
 application domain where bartering takes place is community networks
 where individuals like Alice and Bob contribute resources (for example,
 a segment of the communication link) to a communal
 resource pool to earn the right to access the shared communication and
 computation facilities~\cite{Braem2013}. Also, a barter system
 can be used to barter unused spectrum to save money and contribute to the
 currently pressing issue of spectrum scarcity~\cite{LehrCrowcroft2005}. 
 For example, individuals can barter their unused home 
 bandwidth~\cite{Colin2014}.
   
\item \textbf{services:}  services and skills can also be swapped. Therefore,
 within this category we place conventional
 services whose delivery requires the physical presence of the
 service provider. A good example of service is
 repair work, for example of a plumber~\cite{Ada2009}. 
\end{itemize}

\subsection{Barter models}
\label{barteringmodels}
We anticipate that both ordinary human beings, business 
enterprises and governments will participate in barter 
transactions.  Though in this article  we focus mainly on 
direct (also called peer--to--peer) Individual--to--Individual 
bartering, we expect other models, including Individual--to--Business
bartering and Business--to--Business bartering.

\begin{figure}
\centering
\includegraphics[width=0.55\columnwidth]{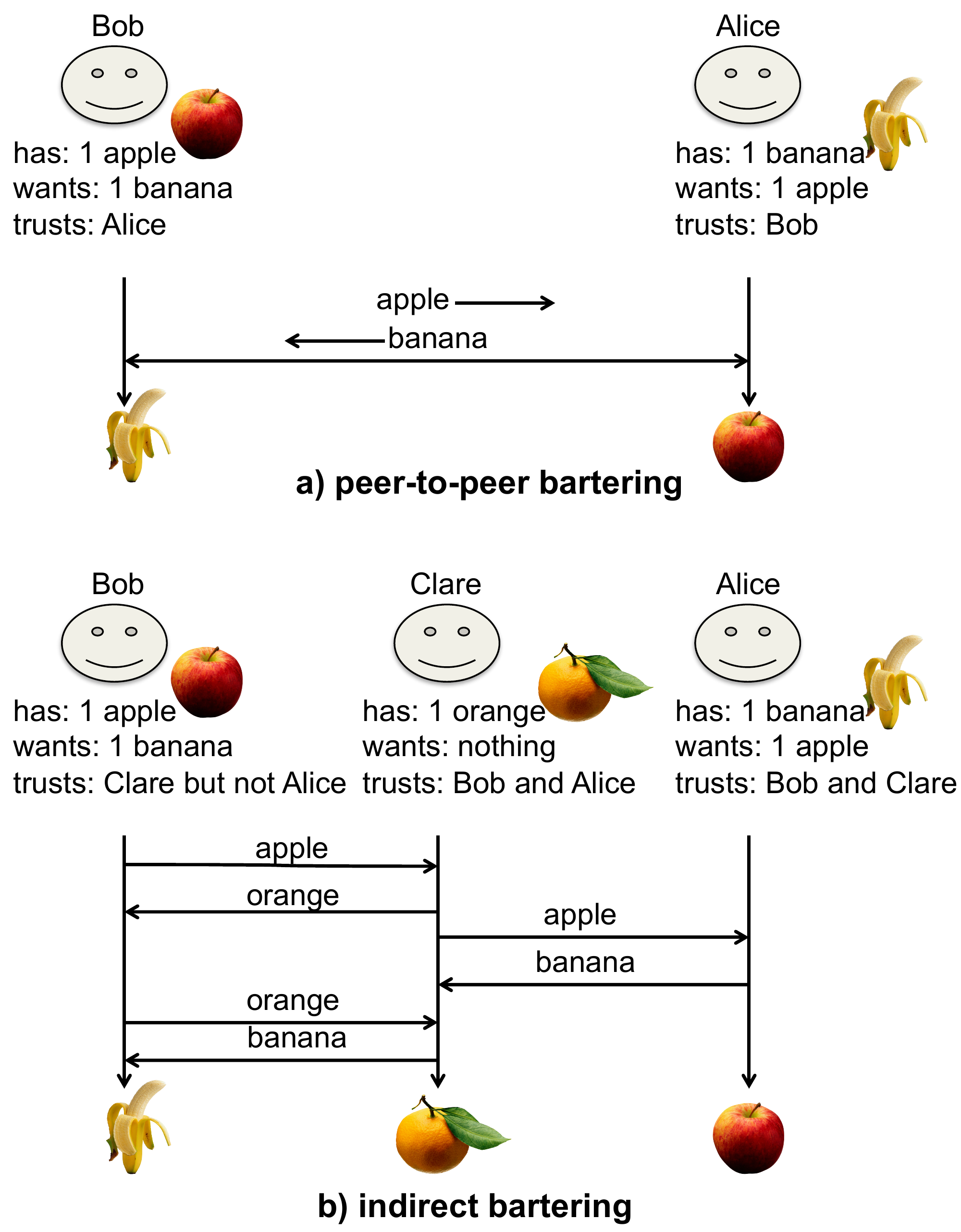}
\caption{Bob, Alice and Clare bartering.}
\label{fig:AliceBobClareBartering}
\end{figure}

In the most simple case, like in the example shown in
Fig.~\ref{fig:aliceandbobbartering}, bartering involves only two 
tangible items and two parties (one item each) that are willing to 
exchange their items simultaneously. 
A similar situation is shown 
in Fig.~\ref{fig:AliceBobClareBartering}--a
where Bob trades with Alice directly: he barters his 
apple in exchange for Alice's banana.
However, by extension of this simple model one can imagine situations of 
arbitrary complexity involving more than two parties where each 
of them owns more than one item.
For example, in the scenario shown in  Fig.~\ref{fig:AliceBobClareBartering}--b, Bob has
an apple, Clare an orange, and Alice a banana.
Imagine that, as shown in the figure, Bob trusts Clare but not Alice, Clare trusts both Bob
and Alice, and Alice trusts Bob and Clare. 
Imagine that Bob is interested in bartering his apple in exchange for 
a banana and that Alice is interested in bartering her banana in return for an 
apple. As shown in the figure, bartering can be conducted only indirectly, 
for example, with Clare's assistance (as opposed to peer--to--peer) 
who operates as a trade exchange (a broker) and for free.
 
One can imagine more elaborated scenarios for example where 
Alice, Bob and Clare have several items to barter and where 
the exchange is $m \geq 1$ for $n \geq 1$  and 
where Clare charges for her brokering service. Notice that this
indirect barter model (see~\cite{RobertKeller1983}) re--introduces 
a mediating party, however, in some applications, the 
flexibility facilitated by Clare can justify her involvement.

The scenario becomes more demanding when more parties
are involved and where one would account for time  constraints and 
exceptions such as potential failures of the parties to deliver 
their items as well as several additional variables and dimensions. 
The complexity becomes more challenging when the bartered items
are complex (for example, electronic or electromechanical
devices) that are expected to deliver a functionality and 
used under the observance of some rules, for example, to
guarantee performance.
A practical barter system would offer means of modelling
tangible objects as digital objects (called tokens) that can be 
negotiated and bartered online.

\subsection{Contribution of bartering to the economy}
We believe that the facilities that decentralised technologies
provide can help bring back the advantages offered by the
ancient barter trade system:

\begin{itemize}
 \item \textbf{peer--to--peer trade:} bartering obviates the
        involvement of fiat cash and consequently of banks in
        trade.  
 \item  \textbf{robustness:} bartering is robust, that is, immune to the
       financial threats that afflict fiat--based trade; this
       valuable characteristic emerges from its simplicity.
 \item \textbf{ad--hoc trade:} with a barter system in place 
        and facilities to locate items and barterers online, ad--hoc
        bartering would flourish.
 \item \textbf{re--distribution of items:} re--introducing of 
       pre--owned items to the economy. To some extent, this is
       already in practice in some countries like the UK
       through garage sales, flea markets, second hand marketplaces, 
       auctions, antique fairs. A barter system would encourage
       the development of this segment of the economy as it
       would facilitate non--monetized exchange at any time.
 \item \textbf{resource sharing:} bartering will facilitate and
       encourage resource sharing as in the example of
       spectrum sharing mentioned above. 
       In the tangible world, it would offer the underpinning technology 
       to support services such as car and accommodation sharing. 

 \item \textbf{reciprocal lending:} bartering can also facilitate 
       the temporary exchange of items where the owners are interested
       only on lending their items, as opposed to parting with
       them permanently. 
\end{itemize}

It is worth noticing that some of these features are
central to Collaborative Consumption and Shared
Economy---two emerging
economic concepts that have the potential to reduce
waste and environmental damage~\cite{Rachel2015,Myriam2016}.

\section{Research challenges}
\label{researchchallenges}
 We suggest a peer--to--peer barter system that offers the 
 necessary services to support the exchange of the items listed in 
 Section~\ref{itemstobarter} and others that are likely to emerge, 
 in a decentralised manner. 
 Such a systems would involve the exchange of items and related documents on a peer--to--peer basis. 
 Barter operations that involve only digital items will be
 completed fully online. That is, Alice and Bob will use their
 devices to exchange the actual digital items, and if they
 wish to, related documents to record 
 evidence of the transaction.
 In barter operations that involve tangible items, Alice
 and Bob will use their devices to exchange related
 documents.
  The exchanged documents might be actual contracts 
  that the participants agree to execute manually or
  automatically after converting them into smart contracts and
  deploying them on blockchains.
 
   In summary, the responsibility of the barter system is to
    offer the necessary services to assist barterers in:
      
 \begin{itemize}
    \item  the exchange of the items, possibly electronically over
           Internet communication channels.
    \item  the exchange of related documents.
    \item  the deployment and execution of contracts that result from
            barter transactions.
    \item  the prevention of the occurrence of disputes where  the
           nature of the items makes dispute prevention  feasible. 
           On the contrary, the barter service needs to provide
           dispute resolution mechanisms.
 \end{itemize}
 
 The barter system should play a passive role, in the sense that 
 it will get involved only when the barter transaction fails to
 complete smoothly. This will happen only occasionally as most of the
 barter transactions will be completed successfully between the
 participants. The implementation of such services will require 
 solutions to several research questions. We will discuss those
 that we consider central.

\subsection{Trust, agreements and legal contracts}
\label{trustandcontracts}
Note that the exchange of items shown in Fig.~\ref{fig:aliceandbobbartering} 
takes place without any documentation. This is
suitable only when the two participants
trust each other completely. This is realistic
when the parties know each other or when the exchange is 
face--to--face and simultaneous  to enable the parties to verify 
that the items are what they expect before accepting them.
However, in the absence of trust, a barter transaction
will involve the exchange of contracts that protect the
participants from potential frauds; for example, a party
delivering the wrong item, delivering too late or not
delivering it at all. We believe that these situations
will be also common.  In Fig.~\ref{fig:aliceandbobbartering}
for example, Alice and Bob are likely to bring and use 
their mobile phones to sign and exchange an agreement  
to document their trade.

We anticipate agreements of different types. At one end of
the spectrum we envisage an informal agreement where the 
participants 
exchange, along with the items, informal documents with 
some basic information about the barter operation,
such as date and names of the participants and descriptions
of the items. The salient feature of these agreement documents
is that they do not need to be enforced as part of the
barter operation; their role is merely prove that the
barter operation took place.

At the other end of the spectrum we envisage situations where
the barter operations are protected by legal contracts
that are signed and exchanged, possibly before the exchange
of the items takes place or during the operation. 
Such contracts will be expressed in a natural language (for 
example, in English) and  stipulate some rights, obligations 
and prohibitions that the two parties are expected to honour 
after remotely signing them. The salient feature of the contracts
is that they need to be enforced either manually or automatically.
The manual enforcement option  is the simplest 
and requires that the parties sign the contract, store it 
safely, for example, on their local disk and retrieve it only 
if one of the parties raises a dispute. 
The automatic enforcement option is more practical 
and more challenging but the technology to implement it is
has been available for decades~\cite{Szabo1997,solaiman2003}. 
We will elaborate on this point below.

\subsubsection{Automatic enforcement of barter contracts}
\label{enforcementbarteringcontracts} 
The executable contract encodes the rules that the 
participants of the barter operation agree to observe. Two examples
of rules that might appear in barter contracts are: 
``Alice and Bob are expected to send their items to 
  each other inmediately'' and 
``Alice is expected to send her item to Bob by Friday midnight and 
Bob is expected to send his within three days after receiving
Alice's''. 
With automatic contract enforcement, failures to observe the rules 
will be detected and signaled by the contract, 
automatically. This requires that the text of the contract
which includes rules like those above,
is converted into executable code (written in a programming language), 
that is capable of monitoring and 
enforcing the contract with little or no human intervention. Automatic 
contract enforcement is far
more demanding due to several associated problems, including the formal 
verification of the smart contract to confirm its logical 
consistency and its deployment. The technology to address these challenges
has been around for decades~\cite{MolinaTSC2011} and is now
more mature.

\subsubsection{Contract deployment}
\label{contractdeployment}
Good alternatives to deploy executable contracts are
decentralised blockchain ledgers. The Bitcoin platform~\cite{BitcoinHome} 
was a significant step in this direction. They demonstrated the use of smart 
contracts (a fancy name for executable contracts) in the implementation of 
Bitcoin. More complex (Turin complete) contracts can be deployed
on the Ethereum ledger~\cite{EthereumHome} and the 
Hyperledger~\cite{HyperledgerFabric}.
 An important hallmark of blockchains is that they offer middleware
 services for implementing decentralised applications such as
 decentralised barter smart contracts. The cost of  
 decentralisation is the execution of consensus protocols 
 among the smart contract instances deployed on the blockchain.
 The consensus protocol inflicts several scalability constraints 
 that are hard to overcome such as transaction
 throughput~\cite{Marko2015,Joseph2016}, 
 consistency, latency~\cite{Decker2016,PeterBailis2013}, 
 and transaction fees. There are reports that  Bitcoin 
 and Ethereum have already experienced average transaction fees of 54.90 
 and 4.15 USD, respectively~\cite{bitinfocharts2018}. 
There are other deployment alternatives that might
help to circumvent these issues. For instance,  in applications
where decentralisation is not a priority, the designer can consider
centralised deployment where the executable contract is
deployed on a centralised trusted third party that offers a contract 
execution environment~\cite{CarlosMolina2019}. 
Also, in applications where centralisation is unacceptable, the
designer should consider  hybrid deployment where the
contract is split into two complimentary pieces: one of
them in deployed on--blockchain and another one
off--blockchain~\cite{CarlosMolinaJan2019}. 
Several hybrid approaches have been reported~\cite{Molina2018,Eberhardt2017,AliDorri2017,Florian2016,Guy2015}.

\subsubsection{Multi--party barter contracts}
\label{multipartybarteringcontracts}
In general, a barter transaction can involve more
than two parties, one or more items each (see Fig.~\ref{fig:AliceBobClareBartering}).
Since collective trust is more  difficult to achieve  than
bilateral, these operations are likely to involve the
exchange of contracts, or more precisely, of multi--party contracts.
Multi--party exchange is more challenging than bilateral
exchange; its complexity depends on the topology of the
interaction (for example, start, ring, graph, etc.).
Some protocols that facilitate multi--party exchange of items have
been reported. However, they do not seen to be mature enough to be
used in practical applications, for instance, they seem to work for 
some classes of items but not for others~\cite{AsokanSchunter1996,Matt1998}.
Regarding contracts, one--to--one oblivious transfer was 
suggested in a pioneering article~\cite{ShimonOded1985} for signing 
contracts between two parties. Multi--party signatures are more
challenging but they have been explored~\cite{Sjouke2015,Payman2006}.

\subsection{Privacy-preserving bartering} 
\label{barteringunderzeroknowledge}
In some situations, barterers might wish (for the sake of privacy and 
business strategies) to conceal some details of their barter
activities from counter parties and third parties. In contrast, in
other situations (for example, to enhance the reputation of the
barterer or the item) they might wish to make these details
public. Examples of details are
the identities of the barterers, offers, counter--offers, agreed
upon values of the items and historical records of 
barter transactions including unsuccessfully negotiated transactions. 
Mechanisms that help to conceal information might be needed
at different stages of the 
barter process, including negotiation and enforcement of 
agreed upon contracts. We believe that Secure Multiparty
Computation (SMC) in combination with zero--knowledge
authentication techniques is a promising
approach to address the problem. 

The idea of SMC is that $N$ parties run a cryptography protocol 
that allows them to compute a function without relying on a 
TTP (Trusted Third Party)
and without revealing to each other the input provided by each party. 
Recent works in barter protocols demonstrate that SMC can be utilized to create privacy-preserving barter protocols. In~\cite{frikken:pbs:2008} the authors demonstrate how to create a privacy-preserving two-party barter system without relying on a TTP. The authors in~\cite{forg:2014,forg:2017} create a privacy-preserving protocol between two parties, where the barter quotes are kept private using homomorphic encryption. The protocol is secure against semi-honest adversaries. W{\"u}ller et al.~\cite{wuller:2016} propose a 
two-party privacy-preserving barter protocol that is secure against 
active attackers and is based on threshold homomorphic cryptosystem.     

Apart from two-party privacy-preserving protocols, there are research 
works that focus on multi-party protocols. W{\"u}ller et al.~\cite{wuller:2017a} suggested the first privacy-preserving multi--party barter protocol where each party creates a quote that includes the offered commodity and the requested commodity alongside the required ranges of the commodity. The quote is kept private from all the other parties. At the end of the barter protocol each party has learned nothing about the other parties' trades. The authors use a threshold homomorphic cryptosystem to realize the multi-party barter protocol.
Another recent proposal for multi-party privacy-preserving barter protocol uses differential privacy to achieve the privacy guarantees required in a barter--based economy~\cite{kannan:2018}. This work requires the presence of a TTP to realize the barter protocol

Regarding zero--knowledge authentication, the idea is that a party can prove 
its identity without leaking information to the challenger. The 
problem of these solutions is the high cost of running the 
zero--knowledge protocol and potential incompatibility
problems when blockchains are involved~\cite{Guy2015}.
These ideas are currently active research topics. Pioneering
work is reported in~\cite{Yao1982} and \cite{Shafi1989}, respectively. 

\subsection{Exchange of tangible items under privacy}
The exchange of tangible items requires the physical
presence of the participants. Consequently, the observance
of privacy is challenging as it conflicts with the need
to identify individuals. There are situations when the 
participants need to be identified to hold them accountable for 
their actions; for example, when the exchange of the items 
is not simultaneous and as result, one of the participants is 
expected to deliver his or her item later, for example, on the 
following weekend.
A potential solution to address the issue is the Proof of Personhood (PoP) 
protocol suggested by Ford~\cite{MariaBorge2017}. The PoP  protocol
provides individuals with pseudonym certificates that protect their 
privacy but makes them accountable for their actions. 

\subsection{Fair Exchange}
\label{fairexchange}
Central to bartering is the operation to exchange the
items and possibly signed contracts (see Section~\ref{trustandcontracts}).
Depending on the properties of the items, the operation
can be executed remotely and entirely automatically
over electronic communication channels.
 
The challenge here is to develop protocols that guarantee fair 
exchange of both, the items and the signed contracts. 
Imagine that Alice owns item $I_A$ and Bob owns item $I_B$, that
the two items can be transferred electronically and that
they have agreed to barter them; to explain the definition of fair
exchange let us assume that the exchange involves only
the items. A fair exchange protocol 
run between Alice and Bob to exchange their items
will guarantees only two alternative outcomes: either success 
or abort. If the exchange completes successfully,
Alice is left in possession of $I_B$ and Bob is left in 
possession of $I_A$ and both assured that the received items
satisfy certain properties and therefore can be accepted
as valid.
Conversely, if the protocol is aborted,  Alice remains
in possession of $I_A$ and Bob remains in possession of 
$I_B$ and neither Alice or Bob is revealed sensitive
information that gives them some advantage. Fair exchange
of contracts is similar: either Alice gets a contract signed
by Bob and Bob gets a contract signed by Alice or the
exchange is aborted.
Fair exchange has been a topic of research interest since 
the 1990s~\cite{Asokan1997} with the success of e--commerce.
Though several protocols have been 
suggested~\cite{Henning2003,Indrajit2002}, they are
mainly based on the use of stateful trusted third parties
that facilitate the exchange and help to solve disputes
when they occur.
The main advantage of using TTP--based protocols is that
they are simple. However, though the 
inclusion of a stateful TTP
is beneficial in some applications, its involvement
brings several economic, privacy and security concerns~\cite{CarlosMolina2019}.
To start with, the TTP charges for its service and might charge extra when 
it is required  to solve disputes. Secondly, the TTP introduces delays. 
More  worryingly, the TTP gathers significant 
amount of sensitive information about the transactions, 
consequently it introduces several risks:
the TTP will delay the progress of the protocol if it
becomes unreachable; also there is a risk that the TTP might 
abscond with the items, loose 
them or collude with one of the participants. 
Additional research is needed to
devise fair exchange protocols that are able to 
exchange items on a peer--to--peer basis; that is, without
the intervention of stateful trusted third parties
and without the occurrence of disputes. Such protocols
would facilitate the exchange of items of any value.

\subsection{Taxation of barter transactions}
\label{taxation} 
One can argue that a  barter transaction involves the
transfer of income, but also of assets in exchange for assets; accordingly, the participants are
subject to tax on the realised gain. The observance of 
this regulation requires that barter transactions are reported to the government tax department for tax purposes.

Taxation of peer--to--peer barter transactions is
extremely challenging because the government currently has no mechanisms to detect the occurrence of barter transactions immediately as in the case for 
currency--based transactions. In a strict peer--to--peer model, immediate report of barter transactions is at the discretion of the participants. However, barter transactions might become detectable by the government 
tax department later, for example, if a participant sells his or her item or barters it through indirect bartering or if a conflict between the participants arises. The deployment of mechanisms to report barter transactions immediately would compromise to a greater or lesser extent the peer--to--peer model.
If this is acceptable, one can for example, delegate the responsibility of reporting the transaction to the trusted mediator that the fair exchange protocol used to exchange the items will inevitably involve~\cite{Henning1999} (see fair exchange bullet of Section~\ref{furtherread}). In this manner, 
the production and notifications of records for taxation purposes is integrated with execution of the transaction. For example, 
the trusted mediator can be programmed to delay the peer--to--peer exchange until it receives a proof that
a transaction in progress has been reported to the 
government.  
 

In the indirect barter model (see Section~\ref{barteringmodels}) the problem is simpler because it assumes the involvement of a
trade exchange which acts as a trusted mediator. A simple solution is to delegate the responsibility of reporting barter transactions to the trade exchange. Once the
transaction is in the government's records, the government can apply the corresponding taxation rules required. 


Tax collection from barter transactions is far more complicated 
than tax collection from currency--based transactions. The difficulty 
is that in a barter transaction, liquidity to pay the tax is absent 
and the valuation of the items traded may be unclear. Several approaches
have been suggested that overcome the difficulties. For 
instance, the taxation can be deferred until the barterers
sell the items. Another possibility is that, the government
waives the taxation of transactions where the barterers are not
making profit or seeking to make profit but contributing to the 
society to ongoing exchange; transactions that involve the exchange of
gardening tools rescued from sheds fall within this
category. These and other ideas are elaborated in~\cite{Elkins2012}. 
This is, however, a difficult bar to integrate into a barter system 
due to some grey area of discerning whether and to what extent 
some or certain barter transactions can be considered to
be the public interest or contributing to society.
Nonetheless much still needs to be done to understand how to 
best integrate tax reporting requirements into barter transactions, 
but also to explore normative solutions
to deal with tax issues, were bartering to become a widespread and 
adopted avenue for conducting business.


\section{Further reading}
\label{furtherread}
  \begin{itemize} 
   \item \textbf{bartering:} to the best of our knowledge, currencyless 
  economy based on 
  barter trade is a  new concept. However, the concepts
  and its practice can be traced back to the early days of
  human civilization. Some historic facts are mentioned
  in~\cite{Intuit2020,Ilana2016}. In~\cite{Ilana2016}
  the authors analyse some disputes about the origins of
  bartering and point out that simultaneous exchange of items 
  was unpractical and therefore less practiced than 
  asynchronous delivery of the items. Cash as legal tender is
  discussed in Section~\ref{currenttradelimitations}. See
  also~\cite{Dror2012}.
  
  \item \textbf{centralisation / re--decentralisation:}
   a motivating discussion of the reason and implications of
   the centralisation of the Internet and the advantages of reversing it,
   is presented in~\cite{JonGareth2020}.
   The same topic, but with a legal and technical view is discussed in~\cite{Enrico2020} and \cite{Eileen2018}, respectively.
   
   \item \textbf{contracts:} smart contracts and decentralised ledgers to deploy 
   them are now widely documented. Satoshi's Bitcoin paper~\cite{Satoshi2008}
  gets all the credits for triggering the now widely
  spread interest on descentralised technologies; complementary,
  the Ethereum yellow paper~\cite{EthreumYellowPaper2018} brought the 
  notion of smart contracts to the attention of the general public.
 Comprehensive and accessible discussions of these topics can be found 
  in~\cite{AndreasAntonopoulos2017} and~\cite{AndreasAntonopoulos2018}, 
 respectively.
  
  The benefits of deploying smart contracts in a decentralised manner are
  several and valuable. That said, let us not forget that decentralisation
  inevitably brings complexity. In situations where centralisation is
  tolerable, the designer might favour simplicity and opt for centralised
  deployment of contracts on trusted third parties. This idea is suggested
  in~\cite{CarlosMolina2019} where the authors argue that full decentralisation
  is extremely difficult to achieve; non--surprisingly, all centralised
  platforms include some elements of centralisation. For instance, to
  be certain about the authenticity of a server with whom a blockchain
  wish to interact, the blockchain would rely on a X.509 certificate issued 
  by centralised authorities. Likewise, the use of oracles
  in some blockchains like Ethereum introduces some degree of centralisation.
  An executable contract is a piece of code written in a 
  programming language; as such, it is likely to include bugs unless 
  counter measures are taken~\cite{Luu2016,Karthikeyan2018,Ahrendt2018}.  
  To reduce the risk of deploying unsound smart
  contracts, designers need tools for conducting formal validation at
  design time by means of model checking complemented by 
  systematic testing of the actual executable code. Model checking is
  aimed at uncovering typical errors that impact contractual
  clauses, such as contradictions, omissions, replications~\cite{MolinaMWSOC2009}. 
 
  \item \textbf{decentralised ledgers:} a decentralised ledger can be implemented
  on the basis of a blockchain data structure, but other data structures can be
  used to avoid or ameliorate scalability and other problems mentioned in 
  Section~\ref{contractdeployment}. Trees and Direct Acycled Graphs (DAG) are 
  examples of data structures that are currently under exploration. A notable example
  if the work being conducted by TODA~\cite{Adam2019}: they have build a distributed ledger on 
  the basis of Merkle tries that does not suffer from
  scalability problems. 
  Another distinguishing feature that makes TODA particularly attractive 
  for implementing barter systems is its inherent notion of 
  exchange of ownership that is implemented in their underlying 
  protocol, as opposed to application level.
  
  \item \textbf{decentralised finance:} the recent rise of Decentralised 
     Finance (DeFi) in the cryptocurrency space is related to our work. To 
     some extent DeFI has some elements of bartering since it enables 
     parties to exchange crypto tokens in a peer--to--peer basis and in a 
     decentralised manner. An early example of a DeFi system is 
     Kybernet~\cite{Loi2017} which facilitates the instant exchange and conversion
     of digital assets like cryptotokens and cryptocurrencies like
     Bitcoins and ETHs.
     More recent examples have proven to be more successful and, unfortunately, 
     more complex. The DAI~\cite{DAIhome2020} token 
     has successfully maintained its ration to USD to about 1:1. Recent research focuses mainly
     on the vulnerabilities that DeFi has inherited from its underlying blockchain construct. 
     A discussion  of flash loan attacks is presented
     in~\cite{gudgeon2020decentralized,qin2020attacking}; the
     argument is that the security hole is caused by
     Ethereum's ``all or none'' transaction execution model.
     
  \item \textbf{privacy:} 
  in~\cite{Fabrice2018}, the authors explore the SMC techniques to provide 
  privacy in cryptocurrency transactions conducted on 
  Hyperledger~\cite{HyperledgerFabric}---a 
  blockchain with support for permissioned access. Practical
  progress has been achieved in this direction. There are libraries 
  for the implementation of APIs to basic SMC, for example, to create, send and evaluate garbled circuits~\cite{Ruiyu2017,YanDavid2011,Pinkas2009}.

  \item \textbf{fair exchange:} a good starting point to understand the relevance
   of fair exchange in bartering and e--commerce is Asokan's 
   PhD Dissertation~\cite{Asokan1997} which can be complemented by the
   concise summary of fair exchange protocols presented in~\cite{Indrajit2002}. Fair exchange impacts 
   payment for items using cryptocurrencies such as bitcoins.
   The lack of a central authority that can offer escrow services 
   makes the exchange of the bitcoin for the item, extremely challenging~\cite{StevenGoldfeder2017}.
   It has been proved that it is impossible to solve
   strong fair exchange without the inclusion of a trusted
   mediators~\cite{Henning1999}. Traditionally, such trusted mediators 
   have been implemented as conventional trusted third
   parties; recently, the availability of trusted hardware embedded in
   commodity devices has opened other possibilities.
\end{itemize}

\section{Conclusions}
\label{conclusions}
We have made a case for a currencyless and cryptocurrencyless 
economy based on bartering. Bartering  could offer an alternative to 
existing currency--centric traditional payment mechanisms, namely,
cash, conventional banking and cryptocurrency. Bartering could
complement and substitute them in a wide variety of instances.

We envisage both fully--fledged business entities and individuals 
bartering with each other online to exchange
digital items, tangible items, services and on--line services. 
In most of these exchanges, transaction contracts (or least some informal documentation) will be used to regulate, outline and govern the 
transactions. Such contracts will be  negotiated, signed, exchanged 
under privacy in a peer--to--peer manner and enforced online after 
deploying them by following centralised, decentralised or hybrid architectures depending on the requirements of the particular exchanges and their parameters.


\begin{table}
 \begin{center}
  {\scriptsize

 \begin{tabular}{|L|L|L|L|L|}
  \cline{2-5}
  \multicolumn{1}{L|}{}  &
  \multicolumn{1}{L|}{cash} & 
                      conventional banking (deposits) & 
                      cryptocurrencies & 
                      bartering    \\
  
  \hline
  \multicolumn{1}{|L|}{\textbf{peer--to--peer}} & yes & no & yes & yes \\
  \hline
  \multicolumn{1}{|L|}{\textbf{privacy}} & yes & no 
                                & \makecell{yes \\ if protection \\ protocols \\ are used} & 
                          \makecell{yes  \\ if protection \\ protocols \\ are used}\\
  \hline
    \multicolumn{1}{|L|}{\textbf{free from transaction fees}} & yes & no & no & yes \\
  \hline
      \multicolumn{1}{|L|}{\textbf{robust to financial uncertainties}} 
                          & no & no & no & yes \\
                          
   \hline
  \multicolumn{1}{|L|}{\textbf{online transactions}} & no & yes & yes & yes\\
  
  \hline
      \multicolumn{1}{|L|}{\textbf{financial services (e.g., credit and escrow)}} 
           & no & yes & no & no \\
  \hline
  \end{tabular}
  } 
 \end{center}
 \caption{Comparison of trade models.}
 \label{table:comparison}
\end{table}

Table~\ref{table:comparison} presents a summary of the main
features that distinguish existing trade models and bartering
as suggested in our work.
We believe that the four models can co--exist independently 
serving their segments of the economy, speculatively, as follows:

\begin{itemize}
 \item \textbf{cash/currency:}  is and will remain as the trade model
       for unbanked individuals who normally conduct 
       small payments in cash that can be carried around. Also, it is a practical
       solution when the banking system is temporarily unavailable and  
       where it is not deployed at all, for example, in remote rural
       communities. People reluctant to leave traces of their
       financial activities will also consider payments in cash.
 \item \textbf{conventional banking:} will be used in situations where the traders  wish to avail of the
 services that banks, as trusted intermediaries, offer.
 Examples of such services include credit, escrow services
 and signed evidence of the occurrence of the transaction.
 \item \textbf{cryptocurrency:}  will be a convenient alternative where the trade partners are reluctant to expose all the details of their transaction or of themselves to the bank and yet, they need to generate an indelible, publicly verifiable and ubiquitous record of the transaction. 
 In practice, such records are needed to
 prove, without revealing unnecessary details, that a payment has been conducted and consequently
 the payer is able to claim the right to access a service.
 \item \textbf{bartering:} will be a convenient trade alternative to save on
  fees and delays caused by mediating parties such as banks and blockchain
  ledgers. Also, bartering will be a convenient alternative to trade under
  strict privacy, in this regard, it is more convenient even than cash--based trade.
\end{itemize}

The trade models can combine their strengths and
support each other. For instance, cryptocurrency deposits 
subject to conditional refunds (for example,
Bitcoin's conditional release of payment) can be used as monetary incentives
at different stages of the barter process to discourage 
misbehaviour~\cite{Ranjit2014,Marcin2016}. For example, if a dishonest party
deliberately delays or interrupts a negotiation process, the cryptocurrency
deposit is refunded to the honest counterpart.

Most pieces of the  technology (blockchain, decentralised smart contracts, 
centralised smart contracts, secure multiparty computation and zero-knowledge 
proofs, fair exchange protocols and so on) that are needed to implement a barter 
system are available but needs further development and integration. 
The main goal of this article is to enable and contribute to the discussion 
and motivate research in 
the direction of barter models and platforms in the everyday lives of 
individuals, businesses and other entities to drive efficiencies around 
cost, speed and privacy.

\section*{Acknowledgment}
Thanks to Rafael Z.~Frantz  (Applied Computing Research Group,
Universidade Regional do Noroeste do Estado Estado do Rio 
Grande do Sul, UNIJUI) for proof--reading and commenting
on a draft of this article.
Ioannis Sfyrakis was supported by the European Research Council (ERC) 
Starting Grant CASCAde (GA n\textsuperscript{o} 716980).
Thanks to Melissa Metzger for the drawing of Fig.~\ref{fig:aliceandbobbartering}.


\bibliographystyle{splncs} 
\bibliography{./biblio/bibliography}
\end{document}